\documentclass[sigconf]{acmart}
\AtBeginDocument{%
  }

\setcopyright{acmlicensed}
\copyrightyear{2018}
\acmYear{2018}
\acmDOI{XXXXXXX.XXXXXXX}
\acmConference[Conference acronym 'XX]{Make sure to enter the correct
  conference title from your rights confirmation email}{June 03--05,
  2018}{Woodstock, NY}
\acmISBN{978-1-4503-XXXX-X/2018/06}

\usepackage{enumitem}
\usepackage{multirow}

\begin{document}

\title{OneSug: The Unified End-to-End Generative Framework for E-commerce Query Suggestion}


\author{Xian Guo}
\authornote{Equal contribution.}
\email{guoxian@kuaishou.com}
\affiliation{%
  \institution{Kuaishou Technology}
  \city{Beijing}
  \state{}
  \country{China}
}

\author{Ben Chen}
\authornotemark[1]
\authornote{Corresponding author.}
\email{chenben03@kuaishou.com}
\affiliation{%
  \institution{Kuaishou Technology}
  \city{Beijing}
  \country{China}
}

\author{Siyuan Wang, Ying Yang}
\email{wangsiyuan12@kuaishou.com}
\affiliation{%
  \institution{Kuaishou Technology}
  \city{Beijing}
  \state{}
  \country{China}
}

\author{Chenyi Lei}
\email{leichenyi@gmail.com}
\affiliation{%
  \institution{Kuaishou Technology}
  \city{Beijing}
  \state{}
  \country{China}
}

\author{Yuqing Ding}
\email{dingyuqing03@kuaishou.com}
\affiliation{%
  \institution{Kuaishou Technology}
  \city{Beijing}
  \state{}
  \country{China}
}

\author{Han Li}
\email{lihan08@kuaishou.com}
\affiliation{%
  \institution{Kuaishou Technology}
  \city{Beijing}
  \state{}
  \country{China}
}

\renewcommand{\shortauthors}{Xian Guo et al.}

\begin{abstract}
Query suggestion plays a crucial role in enhancing user experience in e-commerce search systems by providing relevant query recommendations that align with users' initial input. This module helps users navigate towards personalized preference needs and reduces typing effort, thereby improving search experience. Traditional query suggestion modules usually adopt multi-stage cascading architectures, for making a well trade-off between system response time and business conversion. But they often suffer from inefficiencies and suboptimal performance due to inconsistent optimization objectives across stages. To address these, we propose \textbf{OneSug}, the first end-to-end generative framework for e-commerce query suggestion. OneSug incorporates a prefix2query representation enhancement module to enrich prefixes using semantically and interactively related queries to bridge content and business characteristics, an encoder-decoder generative model that unifies the query suggestion process, and a reward-weighted ranking strategy with behavior-level weights to capture fine-grained user preferences. Extensive evaluations on large-scale industry datasets demonstrate OneSug's ability for effective and efficient query suggestion. Furthermore, OneSug has been successfully deployed for the entire traffic on the e-commerce search engine in Kuaishou platform for over 1 month, with statistically significant improvements in user top click position (-9.33\%), CTR (+2.01\%), Order (+2.04\%), and Revenue (+1.69\%) over the online multi-stage strategy, showing great potential in e-commercial conversion.
\end{abstract}


\maketitle

\begin{figure}[h]
  \centering
  \includegraphics[width=\linewidth]{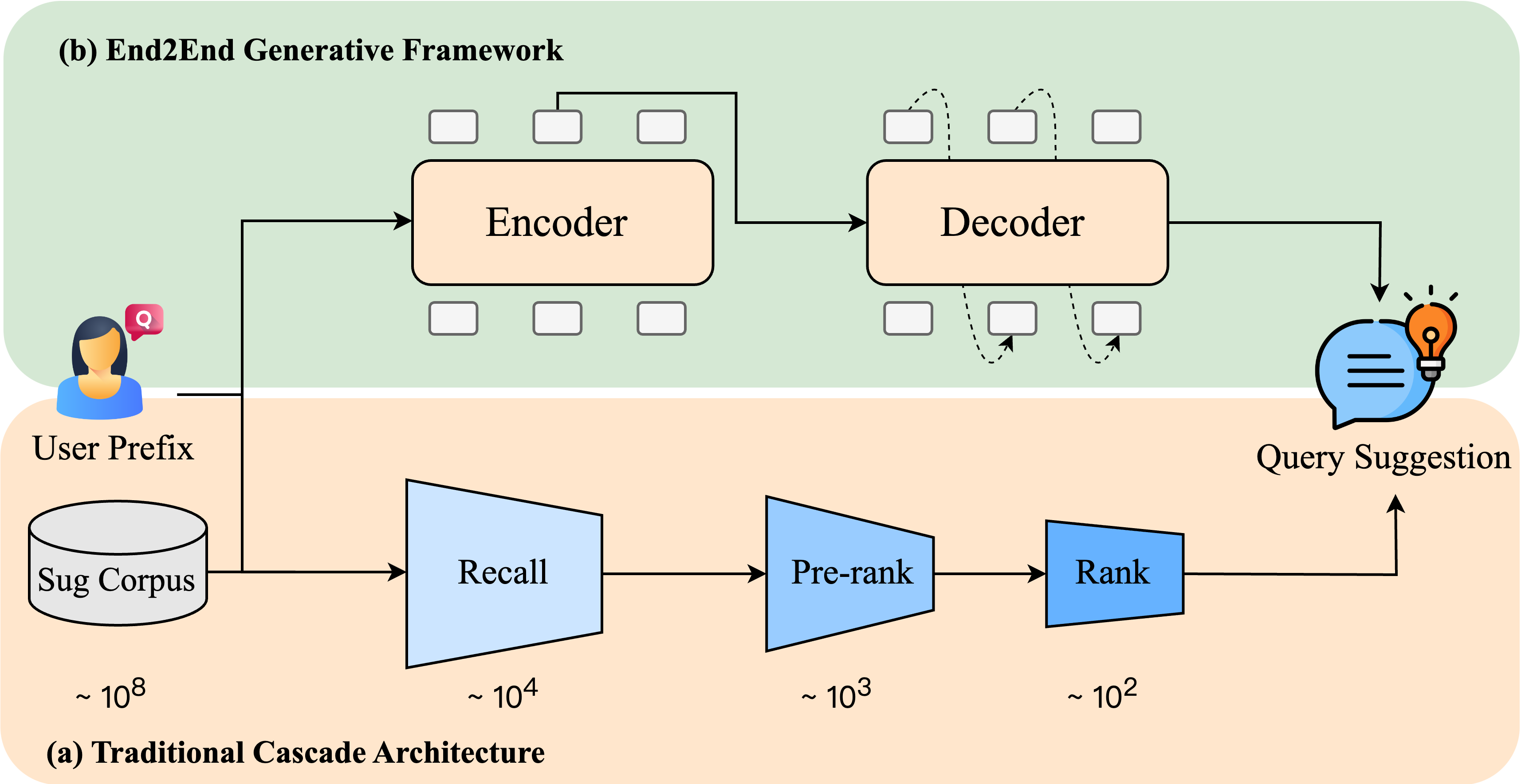}
  \caption{(a) A Traditional Cascade Architecture in Query Suggestion. (b) Our proposed unified end2end generation framework.}
  \label{Figure1}
\end{figure}

\section{Introduction}
Query Suggestion is a fundamental module of modern e-commerce search systems, with the aim of enhancing user experience by suggesting new queries related to the user's initial input. By offering more specific and refined query recommendations, this module assists users in navigating towards their personalized information needs, or providing most frequently searched keywords by users with the same interests, to reduce further word typing thus improving the search efficiency. Typically, consider a user looking for a new smartphone and entering "smartphone" as the query, the search engine might display various brands and models of smartphones, and the user might be interested in a specific brand or function. One useful query suggestion module should provide suggestions such as "smartphones best of 2025" or "smartphone value for money ranking", to assist the users in narrowing down their search to find the item that best meets their needs.

Modern query suggestion modules\cite{ahmad2018multi, chen2020incorporating, lee2024enhanced, li2019click, liu2024mining, mustar2020using} usually adopt past user-system interactions stored in \textit{query logs}, as well as popular search trending words stored in certain pipelines, to refine different search intents. Specifically, to make a balance of efficiency and performance, these modules typically adopt the Multi-stage Cascading Architecture (MCA). An entered prefix in MCA would go through recall \cite{covington2016deep, huang2013learning}, pre-ranking \cite{ma2021towards, wang2020cold}, and ranking \cite{burges2010ranknet, chang2023twin}, finally displaying several queries to the users (Figure~\ref{Figure1} (a)). Each stage is responsible for receiving all queries from the previous stage and selecting the top candidates for the next stage. As shown in Figure~\ref{Figure1} (b), the retrieval will be conducted on all query candidates ($10^8$), while pre-ranking only considers received queries ($10^4$), and ranking stage of $10^2$. 

However, excessive pursuit of the trade-off of system response time and business conversion has resulted in the restricted use of lightweight models in the previous stages (recall, pre-ranking), and complex reasoning can only be carried out in the final ranking stage. This inconsistency makes existing methods have the following limitations: 1) the performance of the previous stage determines the upper bound of the next stage. 2) Heterogeneous modules with different optimization objectives may lead to sub-optimal performance of the overall cascading framework. 3) Existing methods \cite{bar2011context, huang2003relevant, sadikov2010clustering} fail to collect effective queries for unseen prefixes, thus limiting performance in the long-tail sessions. 

The same problems also occur in the cascading framework of e-commerce/video search, recommendation, and advertising engines \cite{ko2022survey, xu2018deep, gharibshah2021user}. To address these, recent works \cite{deng2025onerec, qiu2025one, zheng2025ega, jiang2025large, zhang2025killing, pang2025generative} have devoted to using generative retrieval (GR) methods to replace part or even all of the stage in MCA. Typically, a GR model is trained to directly map items into the ID-based \cite{rajput2023recommender, yang2023auto, zheng2024adapting} or String-based \cite{bevilacqua2022autoregressive, lee2023glen, li2024distillation, pang2025generative} codes, and then generate the identifiers of candidates in an auto-regressive manner. The rich semantic knowledge and powerful reasoning ability of generative models enable them to better capture the fine-grained features of the input, thus surpassing previous methods on recall or ranking. A typical work among them is OneRec \cite{deng2025onerec}, which can replace the entire framework through a session-wise generation approach and an iterative preference alignment module, thereby achieving end-to-end video recommendation. Although it is very promising, this paradigm is not suitable for query suggestion. First, video recommendation is a close-vocabulary task, as its inputs and outputs are both certain videos; while for query suggestion, user inputs and corresponding outputs are open. Secondly, suggestion module must consider the relevance between prefixes and queries. Thus a more fine-grained ranking modeling, other than session-wise, is needed.

In this paper, we propose \textbf{OneSug}, an end-to-end generative framework for E-commerce Query Suggestion. It encompasses:

1) A \textbf{P}refix2query \textbf{R}epresentation \textbf{E}nhancement (PRE) module. Considering that short prefixes (usually only one word) have ambiguous semantics, e.g., “apple” can mean both fruit and company, we attempt to enhance prefix representation using interactively and semantically related queries. We first finetune a representation model with selected <\textit{prefix}, \textit{queries}> pairs to align the content and business characteristics. Then we adopt the RQ-VAE \cite{zeghidour2021soundstream} to generate hierarchical quantitative semantic codes so that each prefix can be enhanced by the queries with the nearly same codes. This module can not only alleviate the insufficient semantic representation of short prefixes, but also combine semantic and business characteristics effectively. In addition, the adoption of RQ-VAE can reduce the giant matching computation during inference, further facilitating the practical deployment of generative models. 

2) An unified encoder-decoder architecture, which takes the prefix, related queries, user's historical interactive queries, and user profile as inputs, and directly outputs queries the user may be interested in. This unified structure is concise, as it effectively avoids suboptimal final results caused by inconsistent optimization goals at the each stages of multi-stage cascading architecture. Thus it can be deployed for end-to-end practical application.

3) A user preference alignment, powered by a \textbf{R}eward-\textbf{W}eighted \textbf{R}anking (RWR) strategy for the generative model. Initially, we categorize user interaction behaviors into six distinct levels and construct nine types of positive and negative sample sequences based on them. Then, we employ Direct Preference Optimization (DPO) \cite{rafailov2023direct} to facilitate the model's learning of preference differences among samples by assigning varying weights according to the level gap. Furthermore, inspired by traditional Click-Through Rate (CTR) models, we extend DPO from contrastive learning with pair-wise data to list-wise one. Subsequently, a hybrid ranking framework, which integrates list-wise and point-wise approaches is proposed, aiming to instilling the effective ranking ability into the generative model and ensure the accuracy of the generated sequence. Unlike straightforward sampling with only selecting the best and worst samples in Iterative Preference Alignment (IPA) \cite{deng2025onerec}, reward-weighted sequence ranking derived from the level gap of user interactive behaviors captures the nuances in user behavior towards different queries more effectively, thus boosting the generative model's capability for concise personalized ranking.

We execute extensive offline evaluations on large-scale industry datasets from the online user search logs, and the significant performance boosts demonstrate the proposed method’s effectiveness for e-commerce query suggestion. Online A/B testings also showcased that it can improve the diversity of query suggestions, and attract more clicks while lowering the query's top click position, ultimately improving business conversions. To the best of our knowledge, it is the first large-scale industrial solution that can provide effective query suggestions via a unified end-to-end generation framework. Moreover, this method has been successfully deployed for
the entire traffic on the e-commerce search engine in Kuaishou platform, with millions of users and serving billions of retrieval PVs, for over 1 month. OneSug yields a 1.82\% decrease in the average input length of prefixes, 9.33\% in user top click position, 2.01\% increase in CTR, 2.04\% in Order, and finally a 1.69\% improvement in total revenue, as well as an average reduction of 43.21\% for system response time, thereby significantly enhancing e-commerce conversion.

\begin{figure}[htp]
  \centering
  \includegraphics[width=\linewidth]{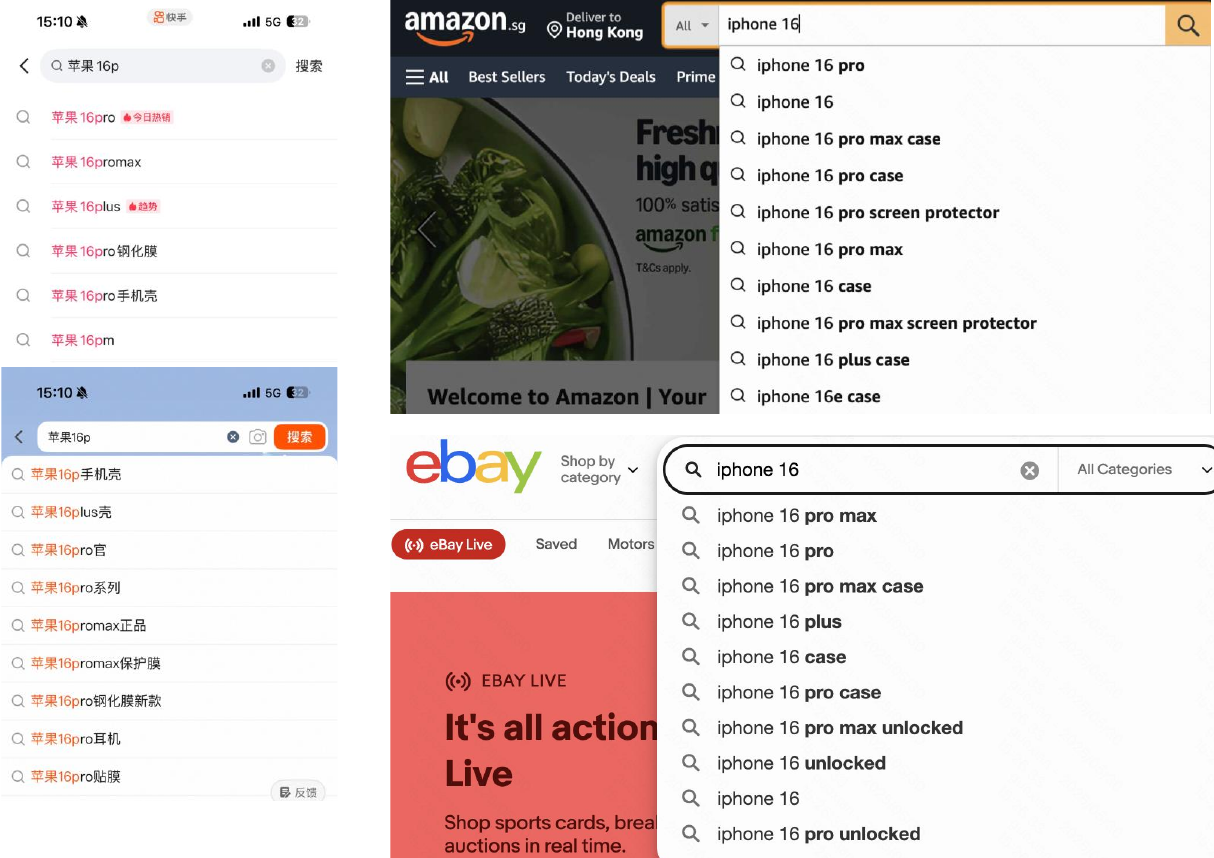}
  \caption{Examples of query suggestion in e-commerce search platforms of Kuaishou, Taobao, Amazon, and eBay.}
  \label{Figure2}
\end{figure}

The main contributions of this work are summarized as follows:
\begin{itemize}[left=0pt]
\item We propose a novel prefix2query representation enhancement module, which uses interactive and semantically related queries to enhance prefix representation, and bridge the gap between the corresponding content and business characteristics.
\item We devise a hybrid reward-weighted ranking strategy derived from the level gap of user interactive behaviors for the generative model. It can distinguish the fine-grained user preference difference to various queries, thus improving the model's personalized ranking capability.
\item We introduce OneSug, an end2end generative framework for e-commerce Query Suggestion. To the best of our knowledge, it is the first industrial solution that 
is comprehensively deployed online with a unified model. Various offline and online A/B tests verify its effectiveness and efficiency, and yield substantial improvements in user click and conversion.
\end{itemize}

\section{Related Works}

\subsection{Generative Retrieval and Recommendation}
In recent years, Generative Retrieval (GR) has garnered significant attention from both academia and industry due to its remarkable performance. This emerging retrieval paradigm, regarding large-scale retrieval as sequence-to-sequence generation tasks, has outperformed traditional ANN-based models such as EBR \cite{huang2020embedding} and RocketQA \cite{qu2020rocketqa}, and spurs increased exploration in the fields of search and recommendation. Notable contributions in this area include Tiger \cite{rajput2023recommender}, DSI \cite{tay2022transformer}, and LC-REC \cite{zheng2024adapting}.

Most GR models serve merely as supplementary recall sources within online systems, thereby overlooking these model's inherent rich semantic and powerful reasoning abilities for the potential uses in (pre-)ranking stages. 
In the area of video recommendation, OneRec \cite{deng2025onerec} first unifies recall, pre-ranking, and ranking within a single generative model, with the assistance of session-wise generation and iterative preference alignment, and achieves a substantial
improvement in practical online metrics.
EGA \cite{zheng2025ega} represents a significant departure from both traditional MCA and existing generative retrieval by introducing a unified framework that holistically models the entire advertising pipeline.
UniROM \cite{qiu2025one} employs a hybrid feature service to efficiently decouple user and advertising features, and RecFormer \cite{li2023text}, a variation of Transformer, to capture both intra- and cross-sequence interactions.
To tackle inconsistent code generation from varying word distributions, GRAM \cite{pang2025generative} boosts retrieval efficiency and pre-ranking accuracy beyond traditional methods. However, its performance is still inferior to online system, owing to the complexities of e-commerce operations.

\begin{figure}[h]
  \centering
  \includegraphics[width=\linewidth]{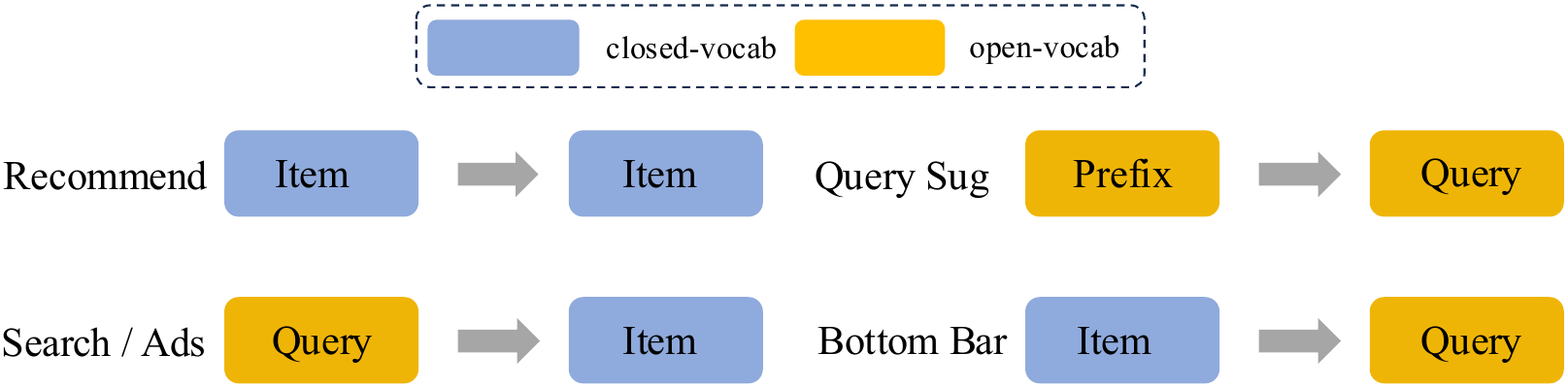}
  \caption{The input and output differences among Recommend, Search/Ads, Query Sug and Bottom Bar.}
  \label{reco_sug_sec_bar}
\end{figure}

While these methods demonstrate appealing performance in the realm of search, recommendation, bottom navigation, and advertising, they are not suitable for online query suggestions. As illustrated in Figure.~\ref{reco_sug_sec_bar}, the inputs and outputs of Query Suggestion are open-vocabulary, which represents a significant departure from both OneRec and GRAM.

\subsection{Query Suggestion}
Previous work on query suggestions uses co-occurrence statistics of prefix and complete query pairs. Of which MPC \cite{bar2011context} suggests top queries that start with the user input prefix through Trie. Other traditional methods adopt users behaviors, term co-occurrence \cite{huang2003relevant}, and keyword clustering \cite{sadikov2010clustering}. These models are efficient in practice, however, fail to generate suggestions for unseen prefixes. 

More recently, neural-networks-based or attention-based methods have paved the way for unseen or longtail prefixs. BART \cite{mustar2020using} is proven to be resistant to noise and suggests more diverse queries than statistics-based models. K-LAMP \cite{baek2024knowledge} utilizes LLM as a generative model and constructs an entity-centric knowledge storage, which is then fed to LLM prompt for personalization. Trie-NLG \cite{maurya2023trie} integrates popularity signals from trie-based methods with personalization signals from previous session queries, thus providing a more robust solution for query suggestion systems. LaD \cite{wang2025personalized} proposes to capture personalized information from both long-term and short-term interests with adaptive detoxification, decreasing the risk of generating toxic content.

Despite significant progress in generative query suggestion models, current approaches are still limited to the recall stage and fail to incorporate ranking signals into the GR model itself. This limitation leads to the loss of critical ranking information during training.

\begin{figure*}[htbp]
    \centering
    \includegraphics[width=\textwidth]{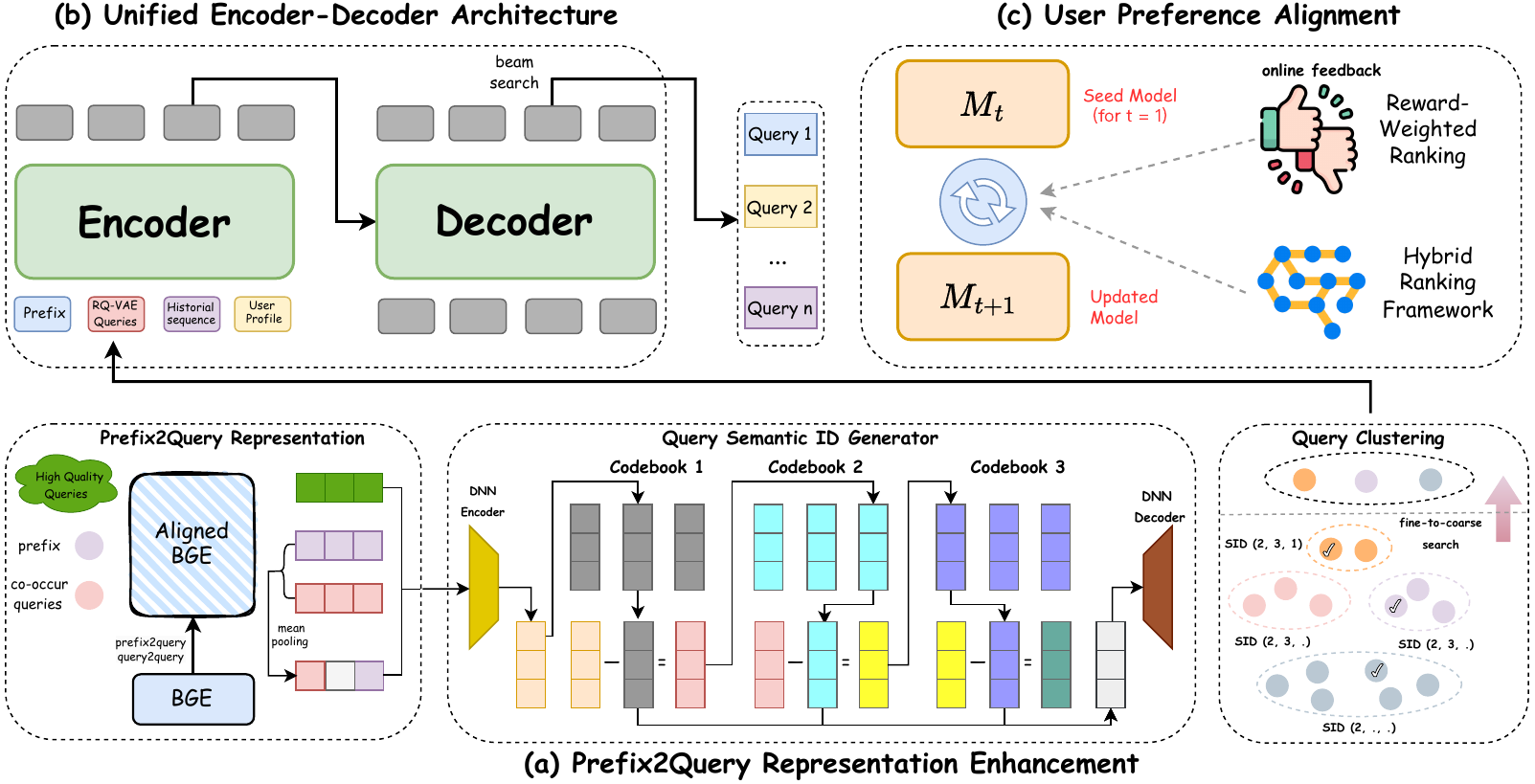}
    \caption{The Framework of OneSug, consists of three components: (a) Prefix2Query Representation Enhancement, provides reference augment for ambiguous prefixes. (b) Unified Encoder-Decoder Architecture, which directly generates queries in an autoregressive manner. (c) User Preference Alignment, which instills ranking ability into DPO training.}
    \label{figure4}
\end{figure*}

\section{Methodology}
In this section, we elaborate on our proposed OneSug in detail, an end-to-end framework for E-commerce Query Suggestion that generates target queries. 
In \S~\ref{Preliminary}, we first introduce the definition of Query Suggestion, and then in \S~\ref{Prefix2Query Representation Enhancement}, a Prefix2Query Representation Enhancement (PRE) module is proposed, which includes the construction of prefix2query representation, query semantic ID generator, and query clustering.
In \S~\ref{Unified Encoder-Decoder Architecture}, we present a unified encoder-decoder architecture that directly generates queries the user may be interested in.
Finally in \S~\ref{User Preference Alignment}, we design the user preference alignment module driven by reward-weighted ranking (RWR) strategy, with the hybrid ranking framework to capture user fine-grained intents.
The framework of OneSug is illustrated in Figure~\ref{figure4}.

\subsection{Preliminary}
\label{Preliminary}
In this section, we introduce the construction of the single-stage end-to-end generative framework for e-commerce query suggestion, from the perspectives of feature engineering. 
The input of OneSug consists of four parts: 1) user input prefix, denoted as \(p\), which is usually an incomplete query, e.g., "smart" corresponding to a potential purchase intention of "smartphone"; 2) the collected prefix2query sequence, which acts as the augment references for \(p\), denoted as \(\mathcal{H}_p = \{q_1^a, q_2^a, \dots, q_m^a\}\), where \(q^a\) represents the query from RQ-VAE and \(m\) is the length of the prefix2query sequence; 3) historical search queries \(\mathcal{H}_u = \{q_1^h, q_2^h, \dots, q_n^h\}\), where \(q^h\) represents the query that the user searched, and \(n\) is the length of the behavior sequence; 4) user profile information \(\mathcal{U}\), which is the crowd portrait fitted by the platform. Then OneSug outputs the corresponding query lists \(\mathcal{Q}\).
OneSug can adopt either encoder-decoer models (e.g. BART \cite{mustar2020using}, mT5 \cite{xue2020mt5}), or the decoder-only models (e.g. Qwen2.5 \cite{qwen2025qwen25technicalreport}) as the foundation. By leveraging the strong instruct-following capabilities and inherent e-commerce knowledge of these generative models, OneSug is expected to generate e-commerce intent queries that meet user's diverse requirements. For the following descriptions, OneSug is denoted as \(\mathcal{M}\).

\subsection{Prefix2Query Representation Enhancement}
\label{Prefix2Query Representation Enhancement}
\subsubsection{Prefix-Query Alignment}
Considering its strong generality across various Chinese NLP tasks, we utilize BGE \cite{bge_embedding} as our initial representation model.
However, its limited knowledge of e-commerce domain necessitates integrating BGE with the real user-perfix-query interactions. Additionally, the fundamental divergence between prefixes and queries leads to misaligned representation spaces, obstructing the further consistent optimization. 
Inspired by QARM \cite{luo2024qarm}, we aim to align representations and infuse latent retrieval knowledge to ensure that BGE can reflect real business characteristics. Specifically, we generate high-quality prefix2query and query2query pairs using existing retrieval models like ItemCF \cite{sarwar2001item} and Swing \cite{yang2020large}, then select the semantically relevant pairs to enhance the performance.
With these high-quality paired data, we train the aligned BGE model through the following procedure:
\begin{equation}
\begin{aligned}
&E_{\text{trigger}} = \text{BGE}\left(T_{\text{trigger}} \right), \\
&E_{\text{target}} = \text{BGE}\left(T_{\text{target}} \right), \\
&\mathcal{L}_{\text{align}} = \text{Batch-Contrastive}\left(E_{\text{trigger}}, \ \ E_{\text{target}} \right).
\end{aligned}
\end{equation}
where $T_{\text{trigger}}$ and $T_{\text{target}}$ denote the original vector of the trigger and target query, $E_{\text{trigger}}$ and $E_{\text{target}}$ represent their embeddings generated from learnable BGE, and $\mathcal{L}_{\text{align}}$ is our alignment training loss. 
Optimizing the loss encourages BGE semantical representations to align with downstream business characteristics.

In the e-commerce query suggestion scenario, the prefixes entered by users are often relatively short (usually only one word), which leads to the ambiguity of prefix semantic representation. To enrich these prefixes with semantically and interactively related queries, we propose the prefix representation enhancement through prefix-query co-occurrence. 
Specifically, for a given prefix \(p\) , we obtain an augmented prefix embedding \textit{$e_p^* \in \mathbb{R}^d$} by using highly-related queries via the aligned BGE model:
\begin{equation}
\begin{aligned}
    &\bar{e_q^c} = \frac{1}{k} \sum_{i=1}^{k} e_{q_i}^c, \\
    &e_p^* = (1 - w) \cdot e_p + w \cdot \bar{e_q^c} \quad \text{where}\ w \in (0, 1),
\end{aligned}
\end{equation}
where \( {e_p} \) is the original prefix embedding, \(e_{q_i}^c\) is the embedding of the \(i\)-th query co-occurred with \(p\), \(\bar{e_q^c}\) is the mean pooling result of these query embeddings, and \(e_p^*\) is the augmented prefix embedding adjusted by weight \(w\). Here, \(w\) for OneSug training is set 0.5.

\subsubsection{Hierarchical Quantitative Semantic ID Generator}
As shown in Figure~\ref{figure4}(a), after aligning prefix representations with high-quality query candidates, we can collect the aligned embeddings of prefixes, as well as those of related queries with similarity values larger than a certain threshold. The next stage is to adopt RQ-VAE \cite{zeghidour2021soundstream} to generate Semantic IDs. The corresponding training objective is shown as follows:
\begin{equation}
\begin{aligned}
&\mathcal{L}(x) := \mathcal{L}_{\text{recon}} + \mathcal{L}_{\text{rqvae}}, \\
&\mathcal{L}_{\text{recon}} := \| x - \widehat{x} \|^2, \\
&\mathcal{L}_{\text{rqvae}} := \sum_{d=0}^{m-1} \| \text{sg}[r_i] - e_{c_i} \|^2 + \beta \| r_i - \text{sg}[e_{c_i}] \|^2,
\end{aligned}
\end{equation}
where \( {x} \), \( \widehat{x} \) is the input of the DNN encoder and the output of the DNN decoder, respectively. $\textit{sg}[\cdot]$ is the stop-gradient operation, \( {r_i} \) is the \textit{i}-th residual embedding, and \( {e_{c_i}} \) is the index of the closest centroid embedding. The intermediate DNN encoder/decoder and the codebooks will be trained in this process.
After extracting the prefix's semantic ID via RQ-VAE, we identify top-\textit{k} most relevant queries through a clustering-based search.
The search process proceeds in a fine-to-coarse manner: first select high-quality queries with the same Semantic ID, then accept those sharing the same codewords.
Additionally, the top-\textit{k} queries will be further screened based on diversity and relevance.
This module can not only alleviate the insufficient semantic representation of short prefixes, but also combine semantic and conversion characteristics effectively. In addition, the adoption of RQ-VAE can reduce the computational complexity of matching during inference, further facilitating the practical deployment of generative models. 

\subsection{Unified Encoder-Decoder Architecture}
\label{Unified Encoder-Decoder Architecture}
Instead of the online multi-stage cascading architecture, OneSug directly outputs the queries that users are most interested in based on the generative architecture using beam search. The output of our OneSug model $\mathcal{M}$ can be formalized as:
\begin{equation}
\mathcal{Q} := \mathcal{M}(p, \mathcal{H}_p, \mathcal{H}_u, \mathcal{U}),
\end{equation}

As illustrated in Figure~\ref{figure4}(b), our model adheres to the transformer-based \cite{mustar2020using} architecture, comprising an encoder that models user historical interactions and a decoder dedicated to query generation.
Specifically, the encoder utilizes stacked multi-head self-attention and feed-forward layers to process \( {p} \), $\mathcal{H}_u$, $\mathcal{H}_p$, and \( \mathcal{U} \), thus produce the encoded historical interaction features $\mathcal{H}$. It can be indicated as $\mathcal{H} = \text{Encoder}(p, \mathcal{H}_u, \mathcal{H}_p, \mathcal{U})$.
The decoder takes the target query's tokens as the inputs and generates the query candidates auto-regressively.
For the unified training, we insert a start token \( t_{\text{[CLS]}} \) at the first place, and a separate token \( t_{\text{[SEP]}} \) between adjacent elements to form the input to the encoder.
\begin{equation}
\begin{split}
x_u = \{ t_{\text{[CLS]}}, p, t_{\text{[SEP]}},\mathcal{H}_p, t_{\text{[SEP]}}, \mathcal{H}_u,t_{\text{[SEP]}},\mathcal{U} \}.
\end{split}
\end{equation}
We utilize the cross-entropy loss for next-token prediction on the token IDs of the target query. 
After a certain amount of training steps on the generation task, we obtain the seed model \( \mathcal{M}_t \).

\begin{table}
  \caption{Online Feedback Of Query Suggestion.}
  \label{tab:freq}
  \begin{tabular}{ccl}
    \toprule
    Online Feedback & Comments\\
    \midrule
    Order& User buy items through specific query\\
    Item Click& User click items through specific query\\
    Click& User click specific query\\
    Show& Specific query show in the pannel\\
    Not Show& Query NOT show, but exists in recall\\
    Rand& Random query in ranking candidates\\
  \bottomrule
\end{tabular}
\end{table}


\subsection{User Preference Alignment}
\label{User Preference Alignment}
\subsubsection{Reward-Weighted Ranking}
Inspired by RLHF's \cite{ouyang2022training, rafailov2023direct} success in NLP, we apply it to our LM-based search system to incorporate diverse user preferences for ranking. 
While OneRec \cite{deng2025onerec} tackles the challenge of annotations with the sparsity of user-item interaction by proposing a personalized multi-task reward model, , our solution instead utilizes feedback from online search systems as a more accessible source of reward signals. Specifically, we categorize the user interactive behaviors in the search system into six distinct levels, as detailed in Table~\ref{tab:freq}. Then we assign the weighted reward score $r(x_u, q) ={\lambda\cdot e^{pi}}$ to each level, where ${\lambda}$ is the base weight with [2.0, 1.5, 1.0, 0.5, 0.2, 0.0] for each level, and ${pi}$ is ratio of each interactive query in the same level. So that the more times a <prefix, query> pair appears at the same level, the greater reward score is assigned.
As a side note, the reason why we did not use CTR as the reward model like OneRec\cite{deng2025onerec} is not only that it requires hundreds or thousands of features to train a high-quality model, but also that it is not conducive to the optimization of the subsequent model's update, due to the high probability of online data deviation.

We take the positive samples among \textit{Order}, \textit{Item Click}, and \textit{Click}, and negative samples among \textit{Show}, \textit{Not Show}, and \textit{Rand}, then construct nine types of sampled <\textit{positive}, \textit{negative}> pairs, e.g. <\textit{Order}, \textit{Show}> and <\textit{Show}, \textit{Rand}>. For each pair the user's preference difference \( rw_{\Delta} \) is computed as: 

\begin{equation}
\begin{aligned}
rw_{\Delta} = \frac{1.0}{r(x_u, q_w) - r(x_u, q_l)},
\end{aligned}
\end{equation}
where \(q_w \) is the win sample,  \(q_l \) is the lose sample. 
Smaller \( rw_{\Delta} \) values encourage model to distinguish the nuances of user interactive behaviors, and also co-occurrence behavior of prefixes and queries.

\subsubsection{Hybrid Ranking Framework}
Relying solely on pair-wise comparisons, Direct Preference Optimization (DPO) \cite{rafailov2023direct} struggles to learn absolute likelihood across different samples, thus leading to an absence of ranking capacity.
With the assistance of RWR, the DPO obtain a certain ranking ability by dynamically adjusting the sample weights.
For generalization, we also propose a target reward margin \( {\delta} > 0 \) to ensure the reward of the chosen sample exceeds the rejected one, at least  \( {\delta} \).
The DPO loss can be optimized as follows:
\begin{equation}
\begin{aligned}
&\mathcal{L}_{\text{pair-wise}} = -\mathbb{E} \Bigg[ \log \sigma \Big( rw_{\Delta} \big(max(0, \hat{r}_\theta(x_u, q_w) - \hat{r}_\theta(x_u, q_l) - \delta)  \big) \Big) \\
& \hspace{150pt} + \alpha \log \pi_\theta(q_w | x_u) \Bigg],\\
\end{aligned}
\end{equation}
where
\begin{equation}
\begin{aligned}
&\hat{r}_\theta(x_u, q_{w/l}) = \beta \log \frac{\pi_\theta(q_{w/l}|x_u)}{\pi_{\text{ref}}(q_{w/l}|x_u)},
\end{aligned}
\end{equation}
\( \hat{r}_\theta(x_u, q_w) \) and \( \hat{r}_\theta(x_u, q_l) \)  is the reward implicitly defined by the language model \( \pi_\theta \) and reference model \( \pi_{\text{ref}} \). To avoid the model solely catering to the reward model at the expense of generation quality, we introduce \( \log \pi_\theta(q_w | x_u) \), which is known as the SFT loss \cite{ouyang2022training}. Furthermore, \( rw_{\Delta} \) is introduced to assign dynamic weight to different samples. 

For a hard negative sampled pair (e.g. <\textit{Click}, \textit{Show}>), the reward difference between the chosen and rejected is small, while \( rw_{\Delta} \) value is large, thus optimization function will enlarge the difference of \( \hat{r}_\theta(x_u, q_w) \) and \( \hat{r}_\theta(x_u, q_l) \) than those of the ordinary pairs (e.g. <\textit{Click}, \textit{Rand}>). So that the nuances of user interactive behaviors to the same prefix can be learned. Finally, with the help of the RWR module, we successfully improve the model's ranking capability.

Conventional DPO is built with the Bradley-Terry \cite{bradley1952rank} preference model, such a training paradigm fails to fully leverage user preference data and overlooks the minor differences of various negative samples for the same prefix, thereby impeding the alignment of LMs with user preferences. 
Inspired by S-DPO \cite{chen_softmax_2024}, we generalize the traditional Plackett-Luce(PL) \cite{plackett1975analysis, debreu1960individual} preference model, which is designed for full relative rankings, to accommodate partial rankings, a more natural fit for recommendation tasks.
In detail, during the preference learning stage, rather than constructing a single <\textit{positive}, \textit{negative}> pairs, we pair input tokens with both positive and multiple negatives to build a more comprehensive preference dataset.
Similarly, we devised a margin loss \cite{boser1992training} for list-wise modeling. By requiring a clear separation larger than \( \delta \), the model becomes less sensitive to minor variations or label noise in the training data.
\begin{equation}
\begin{aligned}
&\mathcal{L}_{\text{list-wise}} = -\mathbb{E} \Biggl[ \log \sigma \bigg( -\log \sum_{q_l \in \mathcal{Q}_l} \exp \Big( rw_{\Delta} \\
&\max \big(0, \hat{r}_\theta(x_u, q_l) - \hat{r}_\theta(x_u, q_w) - \delta \big) \Big) \bigg) + {\alpha \log\pi_\theta\left(q_w | x_u \right)} \Biggr], \\
\end{aligned}
\end{equation}
where \( {\mathcal{Q}_l} \) is set of negative samples.
By combining list-wise preference alignment and the log-likelihood of the preferred sample, we created a new hybrid paradigm for generative ranking.

\section{Experiment}
In this section, we conduct comprehensive evaluations on practical industry datasets offline and rigorous A/B online tests to verify the feasibility of OneSug. Furthermore, we would explore some key questions to facilitate the further research on unified end2end generative model for online serving.

\subsection{Experimental Settings}
\subsubsection{Datasets} 
We extracted the highly reliable user interactive pairs from Kuaishou's online e-commerce logs between February 2025 and March 2025 to facilitate the supervised fine-tuning (SFT) and DPO. It contains about 100 million PVs, and all the following offline and ablation experiments were conducted on the full or part of this data. The collections spanned 32 days, with the first 30 days used for model training and the last 2 days used as the test set.

\subsubsection{Evaluation Metrics}
Since OneSug is designed to make up for the limitations of traditional cascading architectures, here we take into account the recall and ranking performance. We employed HitRate@K and Mean Reciprocal Ranking (MRR) as the evaluation metrics, which are widely used in search and recommendation systems. All data presented were the average values for all tests.

\subsubsection{Baseline Methods}
We compared OneSug with the following two series of representative architectures. The first is Multi-stage Cascading Architecture (MCA). One typical implementation is the combination of multi-stage classic models, just like the comparisons in EGA \cite{zheng2025ega} and UniROM \cite{qiu_one_2025}. Here we adopted BGE for recall, DCN \cite{wang2021dcn} for pre-ranking, and DIN \cite{zhou2018deep}  for ranking. However, it should be noted that real online e-commerce platforms usually use MCA with multiple recalls and complex rankings with at least hundreds of feature combinations. Therefore, using only one model at each stage will actually lead to an unfair comparison of offline performance. In order to more realistically evaluate OneSug, we also used the output results of the real online system (onlineMCA for brief) inputting with the same <\textit{user}, \textit{prefix}> pairs for comparison, even though onlineMCA uses far more features than OneSug. The second is Generative Retrieval Architecture (GRA). As mentioned before, GRA models <\textit{user}, \textit{prefix}, \textit{query}> interactions using transformer-based autoregressive architectures. 
Following the successful implementation of Tiger \cite{rajput2023recommender} and OneRec \cite{deng2025onerec} in video recommendation systems, we construct this baseline by adopting their architectures while replacing semantic IDs with textual inputs and outputs throughout the pipeline.
Furthermore, we implemented OneSug with both encoder-decoder architecture (BART \cite{mustar2020using}, mT5 \cite{xue2020mt5}) and the decoder-only one (Qwen2.5 \cite{qwen2025qwen25technicalreport}).

\subsubsection{Implementation Details}
The selected BGE version is \textit{bge-base-zh-v1.5}.
Considering that the online query suggestion system only displays 16 query candidates for each prefix at a time, the beam search size is set to 32 here to strike a balance between generation quality and latency. The batch size for SFT and DPO is set to 512 and 128, respectively, with the latter being smaller because the list-wise DPO training takes more queries as inputs. For RQ-VAE in the PRE module, the block number \textit{L} of encoder and decoder is 3, the number of codebook layers \textit{C} = 4, and the codebook size \textit{W} of each layer is 512. We take 10 related queries for prefix representation enhancement and 10 historical clicked queries in user logs as context. These hyper-parameters are the most optimal settings validated using the grid search. Some of them will be discussed in the following ablation study.

\subsection{Offline Performance}

\begin{table}[!t]
\centering
\caption{Offline performances of our proposed method OneSug and competitors on industry dataset. The best results are in bold, and the results of the online version are underlined in each column.}
\label{tab:offline}
\begin{tabular}{l|cc|cc} 
\toprule 
\multirow{2}{*}{\textbf{Method}} & \multicolumn{2}{c|}{\textbf{Click}} & \multicolumn{2}{c}{\textbf{Order}} \\
 & HR@16 & MRR & HR@16 & MRR \\
\midrule
MCA & 73.89\% & 39.95\% & 80.71\% & 44.03\%\\
onlineMCA & \underline{78.61\%} & \underline{45.97\%} & \underline{84.55\%} & \underline{51.85\%}\\
\midrule
\midrule
GRA$_{SFT}$  & 73.16\% & 40.06\% & 79.25\% & 44.28\% \\
GRA$_{DPO}$  & 75.50\% & 41.19\% & 81.68\% & 45.30\% \\
\midrule
\midrule
OneSug$_{Bart-B}$  & 82.14\% & 50.55\% & 87.40\% & 56.34\% \\
OneSug$_{Bart-L}$  & 82.84\% & 51.27\% & 88.12\% & 56.80\% \\
OneSug$_{mT5-S}$  & 82.01\% & 50.40\% & 87.26\% & 55.87\% \\
OneSug$_{mT5-B}$  & 83.63\% & 53.01\% & 88.19\% & 57.63\% \\
\midrule
\midrule
OneSug$_{Qwen2.5-0.5B}$  & 85.58\% & 55.34\% & 90.13\% & 60.00\% \\
OneSug$_{Qwen2.5-1.5B}$  & 89.60\% & 60.49\% & 94.95\% & 63.48\% \\
OneSug$_{Qwen2.5-3B}$ & \textbf{93.37\%}  & \textbf{66.31\%} & \textbf{95.13\%} & \textbf{67.40\%}\\
\bottomrule
\end{tabular}
\end{table}

For comprehensive evaluations, we took MCA and onlineMCA as the baselines. As depicted in Table~\ref{tab:offline}, MCA achieves much lower Hitrate and MRR metrics than the online system, as many high-quality queries are eliminated in the initial recall and pre-ranking stages, thus lowering the upper bound of the next stage. OnlineMCA adopts multi-recall and complex (pre-)ranking strategies with hundreds of features, which can alleviate this problem to some extent. However, this operation will amplify the inference cost, significantly increase the system's response time, and degrade the user experience. 

As for the GR models, GRA${_{SFT}}$, GRA${_{DPO}}$ were trained with the input paradigm of <\textit{prefix}, \textit{historical sequence}, \textit{user profile}>. DPO adopted the interactive queries as the positive and sampled the candidates with the lowest CTR score in the same PV as the negative. We can find DPO can enhance the preference modeling, and get a higher performance. But GRA${_{DPO}}$ is still inferior to the online system, as it cannot deeply explore the semantic richness of prefixes, and effectively distinguish the differentiated preferences contained in different user behavior levels. These limitations reduce the effectiveness of GR when applied to industry online engines.

OneSug series achieves the best overall performance by integrating prefix2query representation enhancement, end-to-end generation, and list-wise reward-weighted ranking. Take OneSug$_{Bart-B}$ as an example, it surpasses the performance of both MCA and GRA models. Compared to the onlineMCA, OneSug$_{Bart-B}$ achieves average improvements of 3.19\% and 4.54\% on HR@16 and MRR respectively. While for the larger model OneSug$_{Qwen2.5-3B}$, the improvements can be remarkable at 12.67\% and 17.95\%. These experiments show the promising application of GR models for practical deployment in industrial e-commerce systems.

We also conduct further studies on the effects of different pre-trained architectures and model sizes. For encoder-decoder models, Bart and mT5 of similar size perform similarly, and larger models have better performance. It is more prominent in the decoder-only model, i.e., Qwen-2.5, as the model performance significantly improves when the model size increases from 0.5B to 3B.

\subsection{Ablation Study}

\begin{table}[!t]
\centering
\caption{Ablation Study of OneSug in Offline Evaluations. The best results are in bold.}
\label{tab:ablation}
\begin{tabular}{l|cc|cc}
\toprule 
\multirow{2}{*}{\textbf{Method}} & \multicolumn{2}{c|}{\textbf{Click}} & \multicolumn{2}{c}{\textbf{Order}} \\
 & HR@16 & MRR & HR@16 & MRR\\
\midrule
OneSug$_{list-wise}$ & \textbf{82.14\%} & \textbf{50.55\%} & \textbf{87.40\%} & \textbf{56.34\%} \\
- w/o margin & 81.63\% & 49.70\% & 86.91\% & 55.63\% \\
OneSug$_{pair-wise}$ & 79.39\% & 47.42\% & 85.12\% & 53.01\% \\
- w/o margin & 78.81\% & 46.89\% & 84.62\% & 52.57\% \\
- w/o ${rw_{\Delta}}$ & 77.90\% & 44.41\% & 84.17\% & 49.18\% \\
- w/o ${RWR}$ & 77.28\% & 42.28\% & 82.48\% & 46.66\% \\
- w/o ${PRE\&RWR}$ & 73.16\% & 40.06\% & 79.25\% & 44.28\% \\
\bottomrule
\end{tabular}
\end{table}

Note that we used Bart-B \cite{mustar2020using} as the base pre-trained model for the real online deployment, as it is an economical and effective model and has been online applied in many scenarios in Kuaishou. Here the ablation study and following online testings were conducted on OneSug$_{Bart-B}$. As shown in Table~\ref{tab:ablation}, the combination of prefix2query representation enhancement (PRE) and the list-wise reward-weighted ranking (RWR) module can improve the performance of the GR module with an average of 8.57\% in HR@16, and 11.28\% in MRR. Replacing the list-wise strategy with a pair-wise version, which only inputs one negative sample, causes a considerable degradation of metrics, with reductions of 2.52\% and 3.23\%, respectively. This indicates that diverse negative samples during training help the model quickly learn the differences among various user interactive behavior levels, thereby achieving more concise and effective ranking results. 

We also observed a marked performance improvement with the incremental application of margin loss and \( rw_{\Delta} \), showing the importance of assigning varying weights to RWR module. Furthermore, compared to the results of OneSug$_{pair-wise}$, the removal of RWR decreases the HR@16 by 2.38\%, and MRR by 5.75\%. These all highlight that reward-weighted ranking derived from the level gap of user interactive behaviors can distinguish the nuances in user intent towards different items, thus enhancing the personalized ranking ability of GR models effectively. The removal of the PRE module resulted in noticeable drops for both Hitrate and MRR (-3.68\% and -2.30\% respectively), demonstrating that interactive and semantically related queries are crucial for enhancing the prefix representation from both content and business perspectives.

\begin{figure}[!t]
  \centering
  \includegraphics[width=\linewidth]{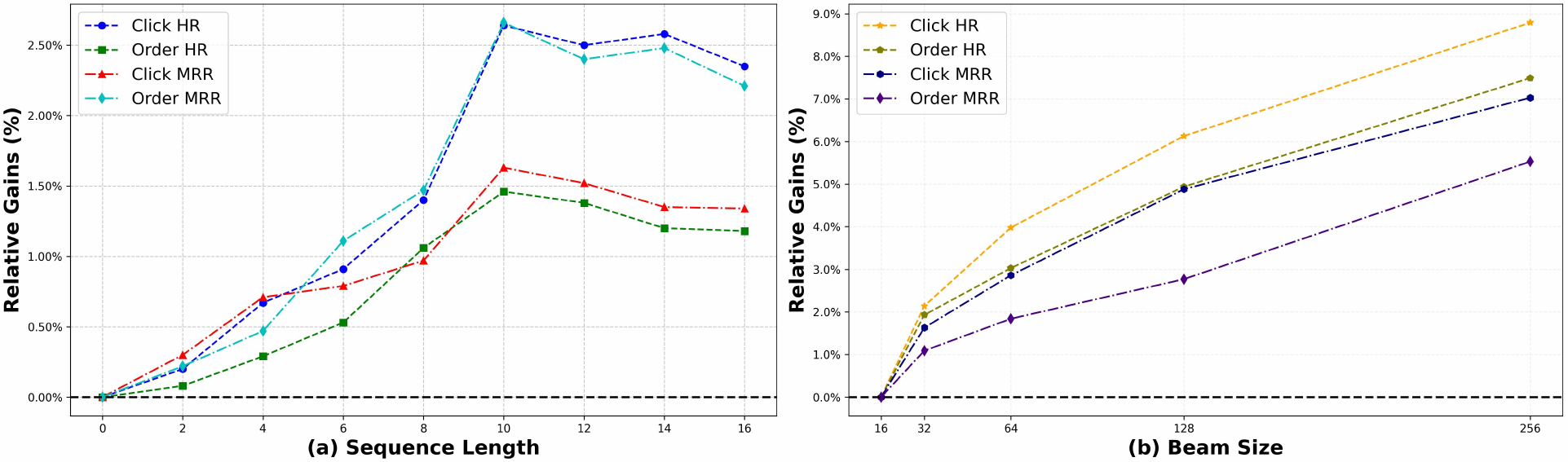}
  \caption{The ablation study of related query sequence length and beam size in inference.}
  \Description{query suggestions.}
  \label{fig:ablation_study}
\end{figure}

Moreover, we also conducted ablation studies on the impact of the related query sequence length in the PRE module and the beam size when inference. As depicted in Figure~\ref{fig:ablation_study}(a), longer does not necessarily mean better, because too long query sequences will introduce more disturbances, which can impair the model's ability to model the user's real intentions and preferences. While the performance of OneSug continues to improve as the beam size increases, from 16 to 256, as plotted in Figure~\ref{fig:ablation_study}(b), but it companies with increasing unacceptable delay. These can induce that moderately longer sequences and bigger beam sizes may lead to more accurate predictions, which requires further research. 

\subsection{Online A/B Testing}
\begin{table}[!t]
\centering
\footnotesize
\caption{Online results for A/B testing.}
\begin{tabular*}{1\linewidth}{@{\extracolsep{\fill}}cccccc@{\extracolsep{\fill}}}
\toprule
\textbf{Method} & \textbf{IPL} & \textbf{TCP} & \textbf{CTR} & \textbf{Order} & \textbf{Revenue} \\
\midrule
OneSug$_{pair-wise}$ & -1.99\% & -9.02\% & +1.78\% & +1.97\% & +1.49\% \\
OneSug$_{list-wise}$ & -1.82\% & -9.33\% & +2.01\% & +2.04\% &  +1.69\%\\
\bottomrule
\end{tabular*}
\label{tab:online}
\end{table}

To verify OneSug’s effectiveness in online applications, we compared
it with onlineMCA in the e-commerce search engine of the Kuaishou platform through rigorous online A/B tests. It takes the short prefix the user entered as input and directly outputs the query candidates, where a query with a higher score would be listed in a more forward exposure position. We assessed the impact of each method on 1) the average input length of the prefix (IPL), 2) user top click position (TCP), 3) click-through rate (CTR), 4) average order volume (Order), and 5) total revenue (Revenue) by comparing performance metrics before and after each implementation. 

As indicated in Table~\ref{tab:online}, OneSug$_{list-wise}$ yields a 1.82\% decrease in the average IPL, 9.33\% in TCP, 2.01\% increase in CTR, 2.04\% in Order, and finally 1.69\% in Revenue, thereby significantly enhancing industry revenue. 
The smaller OneSug$_{pair-wise}$ model demonstrates weaker performance compared to its list-wise counterpart, while the larger OneSug$_{Qwen2.5-0.5B}$ model shows more substantial performance gains.
However, the intensive computational demands limit their practical application in real-time search scenarios with stringent latency requirements. Ultimately, OneSug$_{list-wise}$ has been successfully deployed for the entire traffic on the e-commerce search engine in Kuaishou platform for over 1 month, which serves hundreds of millions of users generating billions of PVs daily.

\begin{figure}[!thp]
  \centering
  \includegraphics[width=\linewidth]{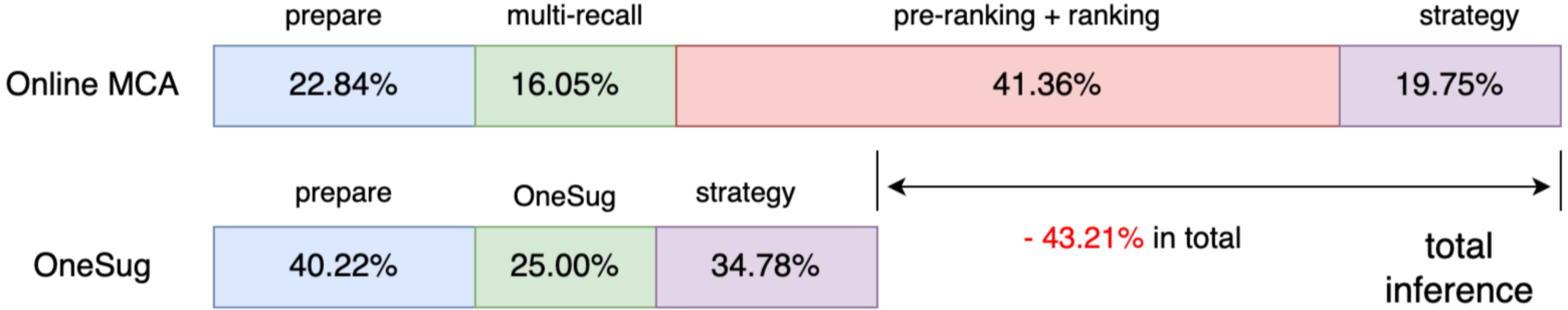}
  \caption{The comparisons of system response time for onlineMCA and OneSug$_{list-wise}$.}
  \Description{query suggestions.}
  \label{fig:RT_time}
\end{figure}

Furthermore, we estimated the average response time of the OneSug model, compared to the online system. As shown in Figure~\ref{fig:RT_time}, OneSug$_{list-wise}$ can replace the multi-retrieval, (pre-)ranking stage, and finally makes an average reduction of 43.21\%. The optimizations of both system response time and business conversion demonstrate that OneSug is an effective and efficient framework for e-commerce query suggestion.

Last but not least, to ascertain the actual impacts on the online search experience, we conducted additional manual evaluations. For each method, we randomly selected 2,000 prefixes and extracted 32,000 prefix-query pairs, ensuring all other variables remained constant. To more comprehensively evaluate the experience performance, we set 1) page good rate - an evaluation indicator for the overall user experience, 2) query good rate - for each query, and 3) full recall rate. Specifically, page good rate requires that all 16 queries under a prefix meet the semantic standards to be a good case. The query good rate evaluates whether each <\textit{prefix}, \textit{query}> meets the requirements. While full recall rate is the ratio of prefixes for which the suggestion system returns 16 query candidates. The outcomes of these assessments are presented in Table~\ref{tab:manual_evaluation}. We can see that OneSug$_{list-wise}$ achieves substantial increases in full recall rate by 8.48\%, page good rate by 11.02\%, and query good rate by 22.51\%. While for OneSug$_{Qwen2.5-0.5B}$, the gain of query good rate can be significant 32.50\%. In conclusion, our proposed method has been demonstrated to substantially improve the user search experience, leading to increased clicks and conversions, and ultimately boosting the industry revenue.

\begin{table}[t]
\centering
\footnotesize
\caption{Manual evaluation results for online experience.}
\begin{tabular*}{1\linewidth}{@{\extracolsep{\fill}}cccc@{\extracolsep{\fill}}}
\toprule
\textbf{Method} & \textbf{full recall rate} & \textbf{page good rate} & \textbf{query good rate} \\
\midrule
OneSug$_{pair-wise}$ & +6.72\% & +9.44\% & +20.49\% \\
OneSug$_{list-wise}$ & +8.48\% & +11.02\% & +22.51\% \\
OneSug$_{Qwen2.5-0.5B}$ & +11.25\% & +18.35\% & +32.50\% \\
\bottomrule
\end{tabular*}
\label{tab:manual_evaluation}
\end{table}

\subsection{Further Analysis}
In this section, we mainly discuss three questions about the online deployment of the end2end generative framework and provide our investigations to facilitate further research.

\begin{figure}[!thp]
  \centering
  \includegraphics[width=\linewidth]{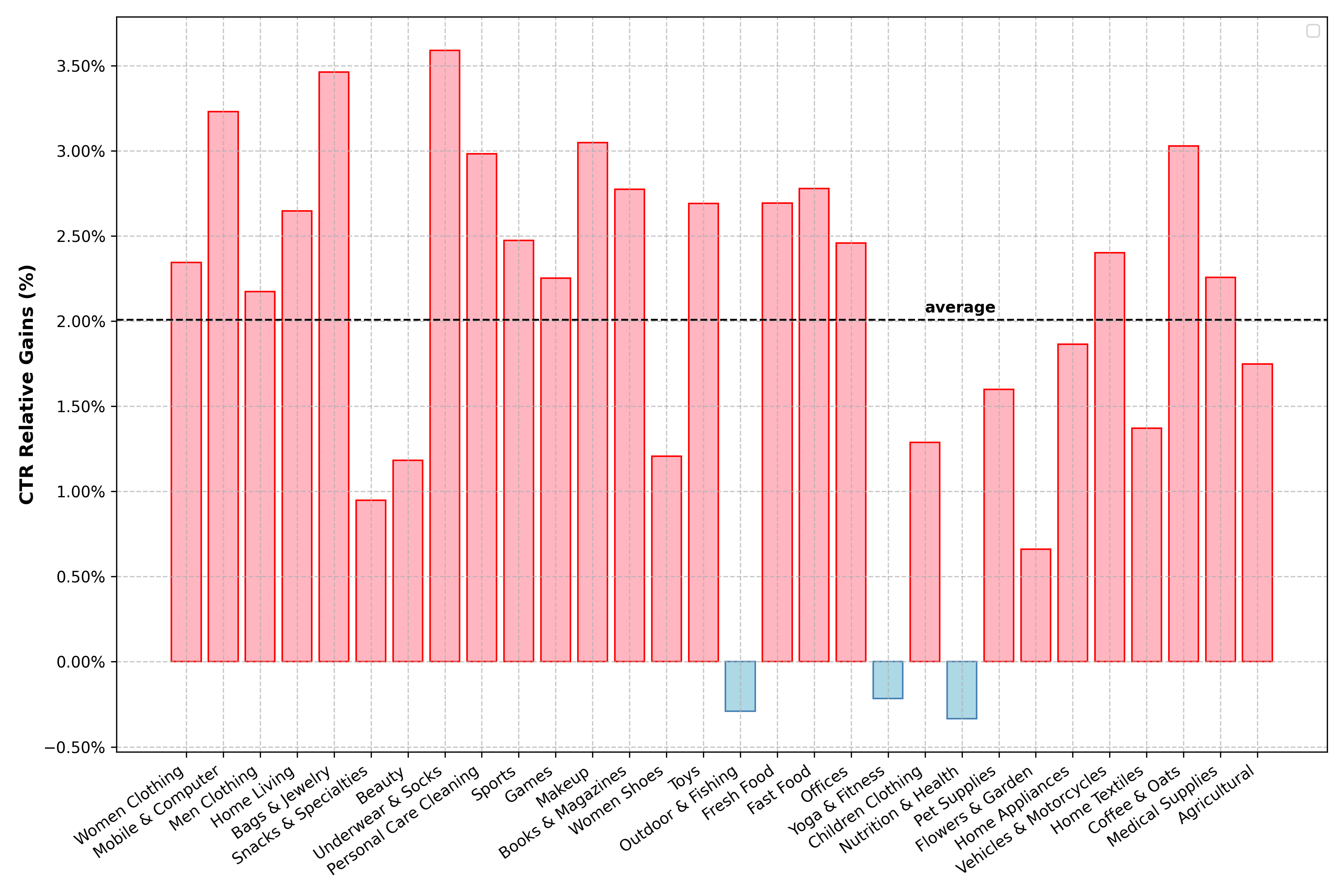}
  \caption{The online CTR relative gains for 30 industries.}
  \Description{query suggestions.}
  \label{fig:gains_by_cate}
\end{figure}

1) \textbf{What are the main aspects of the online gains for the OneSug model?} Here we drilled down from the industries dimension and prefix popularity dimension. As shown in Figure~\ref{fig:gains_by_cate}, We computed the CTR relative gains for 30 industries. 27 of 30 can get the increases with an average of 2.12\%, and the statistical significance (P-value) of these was below 0.05. Three industries were negatively affected, but these values are not significant. Thus almost all industries benefit from the unified modeling optimization, showing the surprising potential for addressing the inconsistent objectives of multi-stage in MCA systems. 

As for the prefix popularity dimension, we divided all prefixes into three categories: top (PV number daily larger than 1,000), middle (larger than 100 and less than 1,000), and long-tail (less than 100). The CTR relative gains for each were listed in Table~\ref{tab:top_middle_tail_query}. Prefixes of all categories are enhanced with either pair-wise or list-wise models, and the gains of the long-tail category are much larger than middle and top categories. These results indicate that the rich semantic and interactive representations inducted by related queries in the PRE module can greatly improve the performance of query suggestion for long-tail queries.

\begin{table}[tp]
\centering
\footnotesize
\caption{Online CTR relative gains for three prefixes categories. Here pair/list indicates pair/list-wise RWR training.}
\begin{tabular*}{1\linewidth}{@{\extracolsep{\fill}}cccc@{\extracolsep{\fill}}}
\toprule
\textbf{Method} & \textbf{Top} & \textbf{Middle} & \textbf{Long-tail} \\
\midrule
OneSug$_{pair-wise}$ & +0.97\% & +1.03\% & +3.26\% \\
OneSug$_{list-wise}$ & +1.15\% & +1.32\% & +3.59\% \\
\bottomrule
\end{tabular*}
\label{tab:top_middle_tail_query}
\end{table}

\begin{figure}[!thp]
  \centering
  \includegraphics[width=\linewidth]{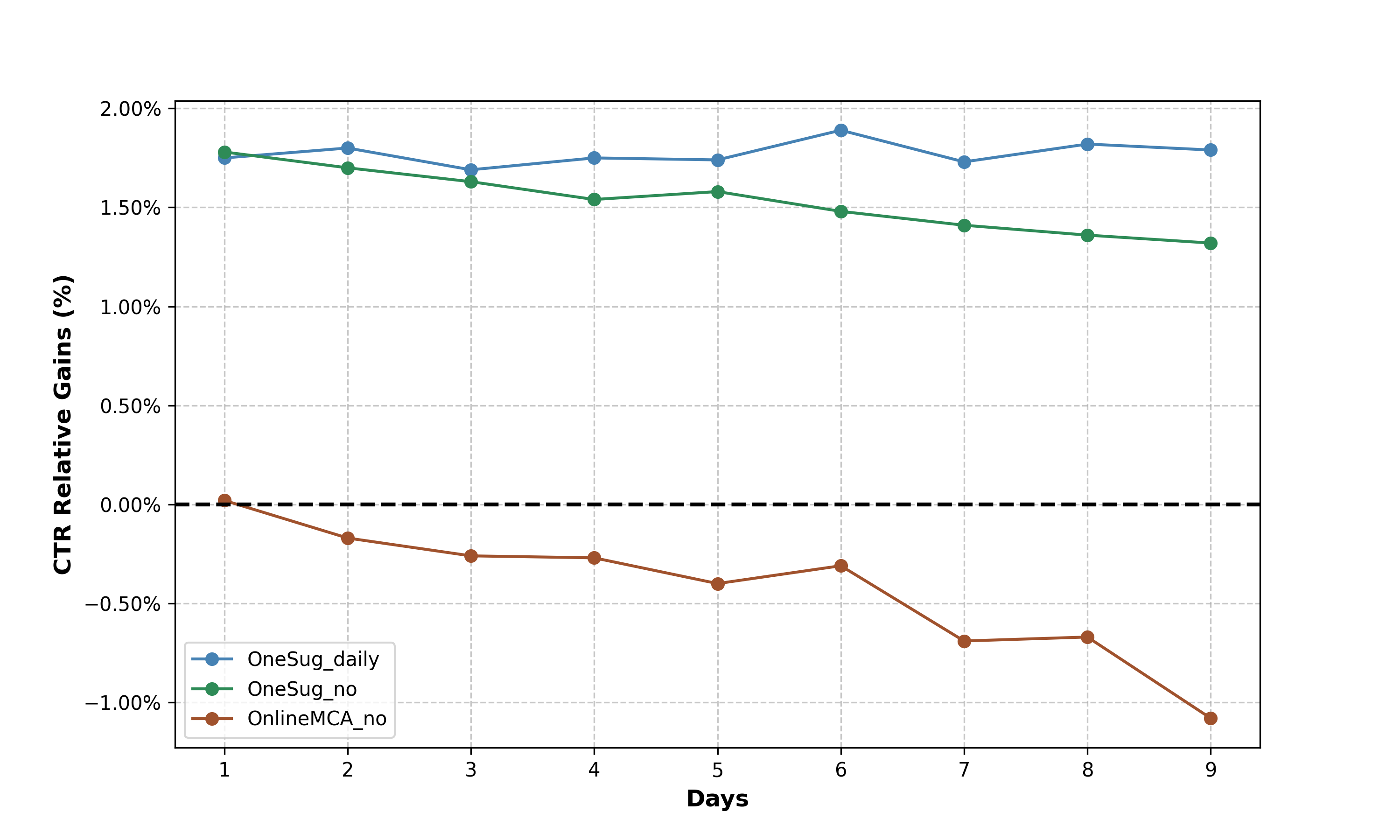}
  \caption{The online CTR relative gains for models with daily updates or not.}
  \Description{query suggestions.}
  \label{fig:gains_by_daily}
\end{figure}

2) \textbf{Does the OneSug model need to be updated regularly?} Models in the (pre-)ranking module of traditional multi-stage cascading architecture often needed to be updated regularly. As plotted in Figure~\ref{fig:gains_by_daily}, here we tested the robustness of onlineMCA and OneSug with no-daily updates (as noted with "${\_no}$"). The baseline is a daily-updated onlineMCA. We can see that both OneSug\_no and onlineMCA\_no are decreased with the day increasing, But the decrease of the OneSug model is much smaller than that of onlineMCA, at -0.6\% compared to -1.1\%. How to keep the generative model updated regularly is a question worth thinking about. Inspired by the way the ranking model is updated, we only use the data from the past three days to update the user preference alignment stage and find that we can maintain effective iterations of the model (noted as "OneSug${\_daily}$" in Figure~\ref{fig:gains_by_daily}) with very low computational costs.

3) \textbf{Do the more features, the better benefits of OneSug?} Similar to RecSys \cite{luo2024qarm}, CTR prediction models for e-commerce suggestion engine are usually trained by the massive real-time user-prefix-query interaction data (i.e., the labels) and a large number of learnable discrete ID-based or String-based features as inputs \cite{naumov2019deep}. These features can be divided into four types:

\begin{itemize}
\item \textbf{ID-based features} contain the User ID, Prefix ID, Category ID, and Session ID. They are usually specified in advance, using hashing or generating randomly, to distinguish different users or PVs. 

\item \textbf{List-wise features} are the query and item lists that the user has clicked in the past, as well as query sequences that are similar to the current prefix, or have the same co-occurrence behavior in the past several days.

\item \textbf{Bucketing features} describes the statistical features of prefixes, users, and queries, such as the number of search PVs for prefixes in the past month, and the number of queries in which users clicked or purchased items.

\item \textbf{Target-aware features} are the cross-attention features computed by the query, item, and prefix sequences.
\end{itemize}

These features have varying effects on generative models. The introduction of ID-based features will significantly reduce the effect of the OneSug model, because the meaningless ID will interfere with the actual semantics of the input sequence. Semantic ID based on RQ-VAE, which is generated by tasks like sequential item prediction and explicit index-language alignment \cite{zheng2024adapting} can effectively alleviate this problem, but it can only be on par with the effects of non-ID-based models. For the open-vocabulary output tasks (as shown in Figure~\ref{reco_sug_sec_bar}), how to efficaciously introduce ID-based features, is a problem worthy of further research.

The target-aware features, like the number of co-occurrences of prefixes and queries, as well as the list-wise and bucketing features, have much less interference with the text input of OneModel. But to better introduce them, a well-crafted prompt is required, and excessive additions will also cause the model effect to decline.

\section{Conclusion}
In this paper, we present OneSug, a pioneering end-to-end generative framework for e-commerce query suggestion that effectively overcomes the limitations of traditional multi-stage cascading architecture. By enhancing prefix representation and employing a unified generative model, OneSug achieves superior semantic understanding and personalized query suggestions. The reward-weighted ranking strategy further refines the model's ability to capture user preferences, leading to improved ranking performance. Extensive offline and online evaluations confirm OneSug's effectiveness in boosting query diversity, click-through rates, and business conversions. Its successful deployment on the Kuaishou platform underscores its practical applicability and potential to significantly enhance industry revenue. OneSug sets a new benchmark for industrial query suggestion solutions, paving the way for future advancements in generative retrieval methods.

\bibliographystyle{ACM-Reference-Format}
\bibliography{sample-base}

\end{document}